\documentclass[9pt,twocolumn,twoside]{osajnl}

\journal{ol} 

\setboolean{shortarticle}{true}

\title{Toward practical weak measurement wavefront sensing: spatial resolution and achromatism}

\author[1,2,3]{Yi Zheng}
\author[1,2,3]{Mu Yang}
\author[1,2,3]{Zheng-Hao Liu}
\author[1,2,3,4]{Jin-Shi Xu}
\author[1,2,3,5]{Chuan-Feng Li}
\author[1,2,3]{Guang-Can Guo}

\affil[1]{CAS Key Laboratory of Quantum Information, University of Science and Technology of China, Hefei 230026, China}
\affil[2]{CAS Center for Excellence in Quantum Information and Quantum Physics, University of Science and Technology of China, Hefei 230026, China}
\affil[3]{Hefei National Laboratory, Hefei 230088, China}
\affil[4]{email: jsxu@ustc.edu.cn}
\affil[5]{email: cfli@ustc.edu.cn}

\begin{abstract}
The weak measurement wavefront sensor detects the phase gradient of light like the Shack–Hartmann sensor does. However, the use of one thin birefringent crystal to displace light beams results in a wavelength-dependent phase difference between the two polarization components, which limits the practical application. Using a Savart plate which consists of two such crystals can compensate for the phase difference and realize achromatic wavefront sensing when combined with an achromatic retarder. We discuss the spatial resolution of the sensor and experimentally reconstruct a wavefront modulated by a pattern. Then we obtain the Zernike coefficients with three different wavelengths before and after modulation. Our work makes this new wavefront sensor more applicable to actual tasks like biomedical imaging.
\end{abstract}

\setboolean{displaycopyright}{true}

\begin{document}

\maketitle

In adaptive optics, wavefront sensors detect the wavefront distortion which is related to the phase of a monochromatic light field. There have been many types of wavefront sensors. The well-known Shack–Hartmann wavefront sensor (SHWS) uses a microlens array to measure the average phase gradient of each aperture \cite{Shack01,Ares:00}. A new type of wavefront sensor, the weak measurement wavefront sensor (WMWS) \cite{Yang2020,Zheng:21} stems from a quantum weak measurement device \cite{Kocsis1170} meant to measure the real part of the weak value \cite{PhysRevLett.60.1351,RevModPhys.86.307} of photonic transverse momentum, which is proportional to the Bohmian velocity of photons in the de Broglie–Bohm theory of quantum mechanics \cite{PhysRev.85.166,PhysRevA.98.042112,Xiao:17,doi:10.1126/sciadv.aav9547}. This quantity also equals the phase gradient of light \cite{RevModPhys.86.307,PhysRevA.98.042112,PhysRevA.100.032125,PhysRevA.104.032221}. Like SHWS, zonal and modal method \cite{Southwell:80,Lane:92} can be applied to reconstruct the phase.

WMWS employs a thin birefringent crystal known as WM (weak measurement) to slightly displace the light component with a given polarization, and a combination of a quarter-wave plate (QWP) and a beam displacer (BD) to separate beams with left- ($L$) and right- ($R$) handed circular polarization. A camera records the intensity distributions of the separated beams, which yield the phase gradient after calculation. Compared with the SHWS, it has a larger spatial resolution but a lower dynamic range \cite{Lee:05} which is inversely proportional to the displacement of WM \cite{Zheng:21}. In our previous works \cite{Yang2020,Zheng:21}, its spatial resolution has not been discussed correctly, and the modal method was only numerically simulated. As far as we know, there are some shortages of this device. From the birefringence theory, an optical path difference (OPD) between the displaced and non-displaced component is inevitably introduced. Slightly tilting WM alters it into an integer multiple of the wavelength ($\lambda$). But when changing $\lambda$, WM needs calibration again. Also, lights of short coherence length become unpolarized after WM and thus useless. A method that two materials are used to compensate for OPD like an achromatic wave plate \cite{Zheng:21} is not recommended now, because in order to restrain its retardance fluctuation, the WM would be very thin, and its displacement would therefore be too short to ensure its sensitivity and resistance from noise.

\begin{figure}[b!]
\centering\includegraphics[width=4cm]{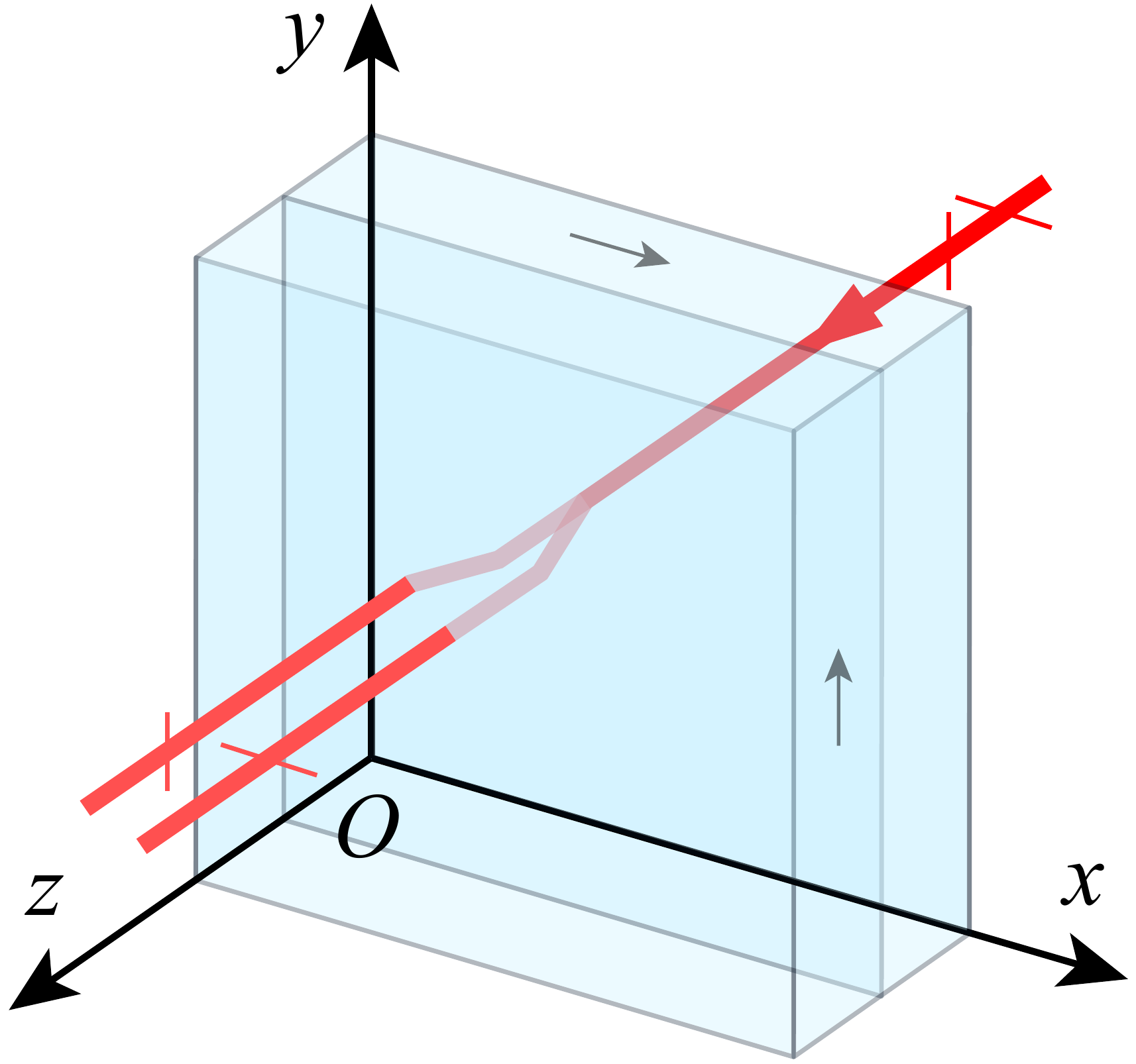}
\caption{The Savart plate. A light beam is incident perpendicularly on the plate. The first birefringent crystal displaces horizontally polarized light toward $+x$ direction, and the second one displaces vertically polarized one toward $+y$ direction.}
\label{p1}
\end{figure}

Here we present another tool which is usually used in fields like polarization interference \cite{Richartz:48,Zhang:11}, the Savart plate (SP), as shown in Fig.~\ref{p1}, consisting of two identical birefringent crystals which are rotated by $90^\circ$ with respect to each other. One of them displaces horizontally ($H$) polarized light toward $+x$ direction, and the other displaces vertically ($V$) polarized one toward $+y$ direction. The OPD is eliminated as both the polarization components go through a displacement and a stationary process. Note that $l$ is still dependent on $\lambda$ due to dispersion, which can be solved by using two materials to make SP itself roughly achromatic \cite{Mu:14}. Then, the QWP should be replaced by an achromatic light retarder. But as we will see, the flatness of the retardance curve is important, and finding a perfect retarder is not easy. The BD can be replaced by a PBS, or we can use a polarization camera, to make this device suitable for a white light within a certain $\lambda$ range. In this Letter, we explain the principle of achromatic WMWS, discuss its higher spatial resolution than a common SHWS, reconstruct the wavefront modulated by a phase pattern, and obtain the Zernike coefficients of wavefronts modulated by a combination of Zernike polynomials of three wavelengths.

The phase gradient of light $\textbf{\textit{k}}=(k_x,k_y)=\nabla\phi=\nabla\arg\psi=\operatorname{Im}[(\nabla\psi)/\psi]$ is related to the complex amplitude denoted as $\psi(x,y)=A(x,y)\exp\left[\textrm{i}\phi(x,y)\right]$ like a quantum wavefunction \cite{Lundeen2011}, though it can be explained in purely classical optics \cite{Zheng:21}. The displacement of a single WM $l$ can be calculated from its thickness and optical axis orientation using the birefringence theory. When measuring $k_x$, the first WM of SP displaces diagonally ($D$) polarized light toward $45^\circ$ direction, and the wavefunction becomes $\psi_\nearrow(x,y)=\psi(x-l_\textrm{t}/2,y-l_\textrm{t}/2)$, where $l_\textrm{t}=\sqrt{2}l$. The second WM displaces anti-diagonally ($A$) polarized one toward $135^\circ$ direction, i.e.~$\psi(x,y)\rightarrow\psi_\nwarrow(x,y)=\psi(x+l_\textrm{t}/2,y-l_\textrm{t}/2)$. If we measure $k_y$, the whole SP is rotated counterclockwise (facing the light) by $90^\circ$. Here we use measuring $k_x$ as an example to describe its principle. Using a polarizer like a PBS, a light beam with $H$ polarization is prepared. Using Dirac's notation, we denote the initial state as $\left|\varPsi\right\rangle=\left|H\right\rangle\left|\psi\right\rangle=\left(\left|D\right\rangle+\left|A\right\rangle\right)\left|\psi\right\rangle/\sqrt{2}$. After passing the SP, the state becomes $\left|\varPsi'\right\rangle=\left(\left|D\right\rangle\left|\psi_\nearrow\right\rangle+\left|A\right\rangle\left|\psi_\nwarrow\right\rangle\right)/\sqrt{2}$. Then $L$ and $R$ polarization component are separated. Letting $\psi_{x,L/R}(x,y)=\left\langle L/R\right|\left\langle x,y|\varPsi'\right\rangle$, where $\left|L/R\right\rangle=\left(\left|H\right\rangle\pm\textrm{i}\left|V\right\rangle\right)/\sqrt{2}$, the intensity distributions at the camera are
\begin{align}\label{eq4}
I_{x,L/R}&\propto\left|\psi_{x,L/R}(x,y)\right|^2\nonumber\\
&=\frac{1}{4}\left[A_\nearrow^2+A_\nwarrow^2\mp2A_\nearrow A_\nwarrow\sin\left(\phi_\nwarrow-\phi_\nearrow\right)\right]\nonumber\\
&\approx\frac{A(x,y-l_\textrm{t}/2)^2}{2}\left\{1\mp\sin\left[k_x\left(x,y-\frac{l_\textrm{t}}{2}\right)l_\textrm{t}\right]\right\},
\end{align}
where approximations $A_\nearrow+A_\nwarrow\approx2A(x,y-l_\textrm{t}/2)$ and $\phi_\nwarrow-\phi_\nearrow\approx k_x(x,y-l_\textrm{t}/2)l_\textrm{t}$ have been used. Then, we have
\begin{align}\label{eq6}
k_x\left(x,y-\frac{l_\textrm{t}}{2}\right)\approx\frac{1}{l_\textrm{t}}\arcsin\frac{I_{x,R}(x,y)-I_{x,L}(x,y)}{I_{x,R}(x,y)+I_{x,L}(x,y)},\nonumber\\
k_y\left(x+\frac{l_\textrm{t}}{2},y\right)\approx\frac{1}{l_\textrm{t}}\arcsin\frac{I_{y,L}(x,y)-I_{y,R}(x,y)}{I_{y,R}(x,y)+I_{y,L}(x,y)},
\end{align}
which yields the phase difference exactly if the amplitude $A$ is constant. The intensity of light can be approximated by $I(x,y-l/\sqrt{2})\approx I_{x,L}(x,y)+I_{x,R}(x,y)$. The maximum detectable $\left|k_x\right|$ and $\left|k_y\right|$ is $\pi/(2l_\textrm{t})$, which is the dynamic range of WMWS, often lower than an SHWS \cite{Zheng:21}, while the sensitivity is improved like the pyramid wavefront sensor \cite{Pyramid,Guthery:21}. WMWS requires the polarization of the incoming light to be uniform, which can be unpolarized but should not possess spatial structure like vector beams \cite{Rosales_Guzm_n_2018}. For an unpolarized light, the reflected path with $V$ polarization can be utilized to measure $k_y$ by another WMWS to realize real-time sensing.

If there is a phase difference $\delta$ ($|\delta|\leq\pi/2$) between the two polarization components, the polarization state of the photons with constant spatial wavefunction $\psi$ becomes $\left|\delta\right\rangle=\left(\exp(\textrm{i}\delta)\left|D\right\rangle+\left|A\right\rangle\right)/\sqrt{2}$, and the intensity of $L$ and $R$ component at the camera will be proportional to $\left|\left\langle L/R|\delta\right\rangle\right|^2=(1\pm\sin\delta)/2$. From \eqref{eq6}, $k_x$ ($k_y$) will be $\mp\delta/l_\textrm{t}$ rather than zero although the light is perpendicularly incident. As for achromatic quarter-wave retarders, only a superachromatic QWP made of at least three materials or a Fresnel rhomb \cite{Serpenguzel:99} satisfy the requirement that the retardance curve is flat enough. We experimentally tested the Fresnel rhomb but found that the wavefront is significantly distorted, possibly because its surfaces are not perfectly flat. So, our experiments use a compromising method: changing the QWP each time we switch $\lambda$.

\begin{figure}[b!]
\centering\includegraphics[width=8.8cm]{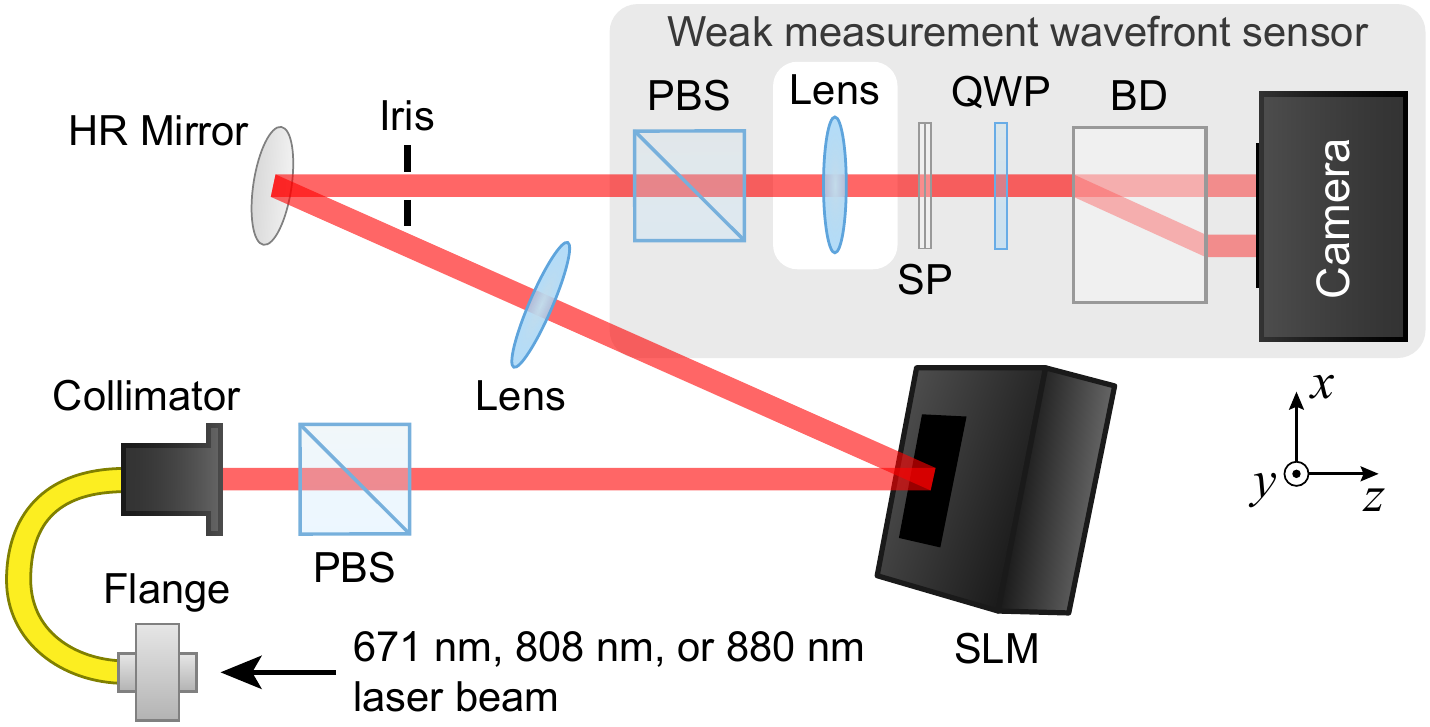}
\caption{The experimental setup. A single-mode fiber (not shown) transmitting a laser beam with a wavelength of 671 nm, 808 nm, or 880 nm is plugged into the flange. After passing through the collimator and a polarization beam splitter (PBS), its phase is modulated by a spatial light modulator (SLM). A pair of Fourier lenses is used to image the optical field onto the camera. An iris is placed at the spectrum plane. The weak measurement wavefront sensor (WMWS) consisting of a PBS, a Savart plate (SP), a quarter-wave plate (QWP), a beam displacer (BD), and a CMOS camera detects the phase gradient of the light. HR: Highly reflective.}
\label{p2}
\end{figure}

The experimental setup shown in Fig.~\ref{p2} consists of two parts, input light beam preparation and wavefront sensing. A single-mode fiber transmitting laser beam with wavelength 671 nm, 808 nm, or 880 nm can be plugged into the flange. The beam comes to free space as a Gaussian beam via a collimator and passes through a polarizing beam splitter (PBS) to prepare $H$ polarization. It is then reflected by a spatial light modulator (SLM), where phase patterns are added. The WMWS consists of another PBS (to ensure the polarization of light), an SP, a QWP (to be switched), a BD, and a CMOS camera whose pixel width is 5.04 \textmu m. Considering the diffraction of light, it detects the phase of the light freely propagating to the camera. We use a $4f$ imaging system to measure the light field at the SLM \cite{Kocsis1170,PhysRevA.104.032221,Zheng:21}. The distance between the second lens and the camera should be larger than the focal length of the Fourier lenses as light beams diffracts less in media like PBS and BD. An iris is placed at the spectrum plane to select the first-order diffracted light. The phase loaded on SLM includes the aberration we need to impose and a grating whose period is proportional to $\lambda$ to ensure the reflected light is at the same angle for different wavelengths. The SP is formed by putting two identically manufactured calcite plates together. The angles between their surfaces and optical axes are $45^\circ$. It is fixed after calibrating it with respect to 880 nm light, i.e.~tilting it to ensure the maximum intensities of the $L$ and $R$ component are the same and the light coming out of SP is not at $V$ polarization. The reason is that the displacement leads to an OPD \cite{Zheng:21} and thus an intensity difference if the SP is not perpendicular to the reference light. If the OPD equals $n\lambda/2$ ($n$ is a non-zero integer), the polarization will be switched. If it equals $n\lambda$, the result may not be affected for this $\lambda$, but there will be a phase difference for other $\lambda$ values.

\begin{figure}[b!]
\centering\includegraphics[width=8.8cm]{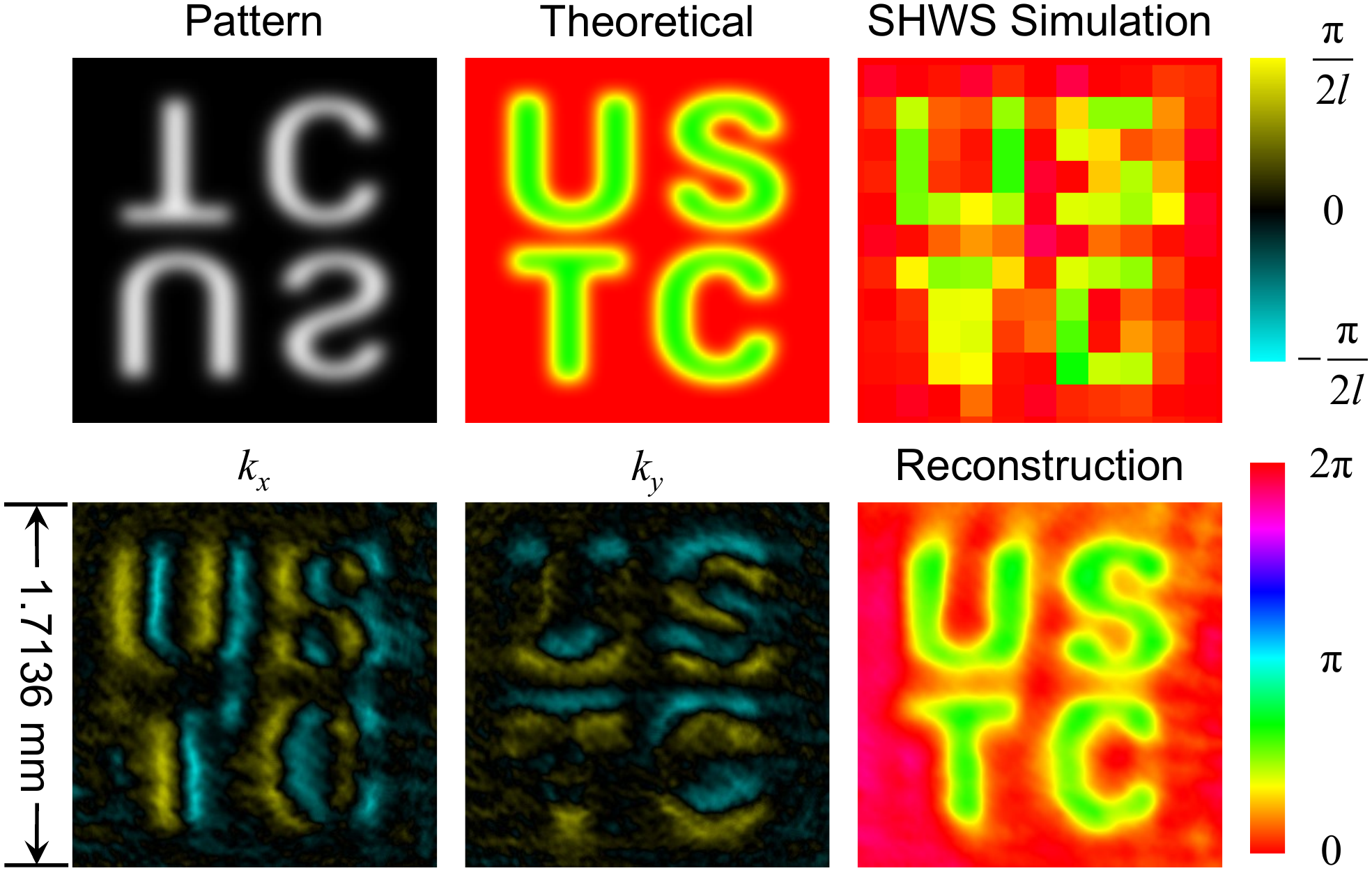}
\caption{The reconstruction of a phase pattern. The square phase pattern “USTC” (Pattern) is loaded on the spatial light modulator. The phase distribution directly from the pattern is denoted as Theoretical, and a numerically simulated phase reconstruction of a Shack–Hartmann wavefront sensor with lenslet size 150 \textmu m is denoted as SHWS Simulation. The measured $k_x$, $k_y$ using the weak measurement wavefront sensor, and the reconstructed phase distribution using the zonal method are denoted as $k_x$, $k_y$ and Reconstruction respectively.}
\label{p3}
\end{figure}

The spatial resolution of WMWS seems to be directly related to the camera, and wavefront sensing of a light beam scattered by a diffuser has been demonstrated \cite{Yang2020}. In fact, this property is determined by $l_\textrm{t}$. For example, a wavefront with its phase $\phi(x,y)=c\sin(2\pi x/l_\textrm{t})$ where $c$ is a constant will lead to $k_x=k_y=0$. From the Nyquist–Shannon sampling theorem \cite{sampling}, the smallest period of phase function allowed is $2l_\textrm{t}$. We just use $l_\textrm{t}$ to describe its spatial resolution, while we use the size of each lenslet to describe that of SHWS. In our experiment, $l_\textrm{t}\approx 51.5$ \textmu m for 880 nm light, which is smaller than the lenslet size of a common SHWS, for example, 150 \textmu m. Nevertheless, a high-resolution camera provides more details to reduce the error. Experimentally, we loaded a phase pattern of letters “USTC” which are blurred to avoid sharp phase changes exceeding the dynamic range, and reconstructed the wavefront shown in Fig.~\ref{p3} using the WMWS with the zonal method, as patterns are not simple aberrations. The root-mean-square error (RMSE) from the reconstructed phase and the pattern itself is $0.0469\lambda$, which might be smaller if we were able to compare it with the inaccessible actual wavefront after the $4f$ system with an iris. We also performed a numerical simulation of the wavefront detected and reconstructed by the SHWS mentioned above, where the letter “S” is obscure. Note that for WMWS, the dynamic range increases with the spatial resolution, contrary to the SHWS with a fixed focal length.

\begin{figure*}[t]
\centering\includegraphics[width=18.4cm]{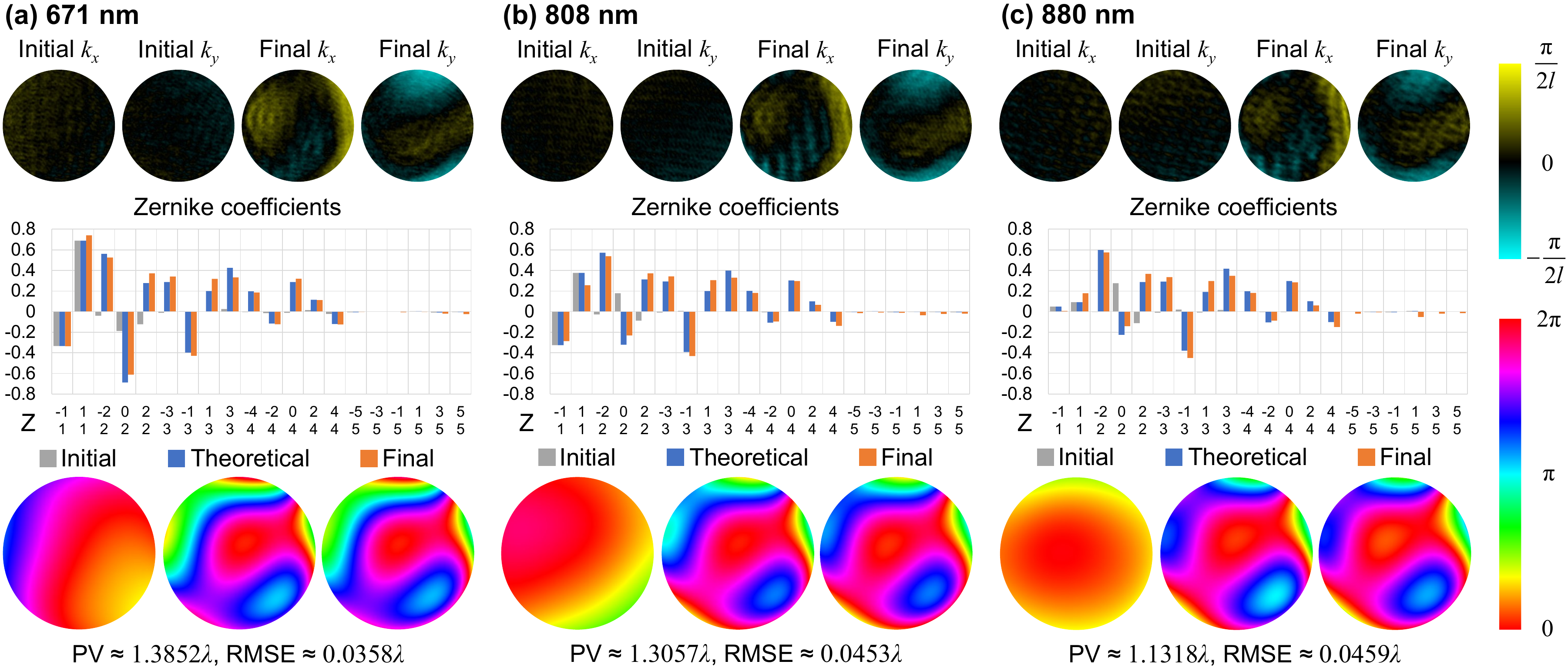}
\caption{The $k_x,k_y$ distribution, Zernike coefficients from $n=1$ to $n=5$, phase distributions of (a) 671 nm, (b) 808 nm, and (c) 880 nm light when the spatial light modulator (SLM) is loaded with only the grating (Initial) or together with the phase aberration $0.6Z_2^{-2}-0.5Z_2^0+0.4Z_2^2+0.3Z_3^{-3}-0.4Z_3^{-1}+0.2Z_3^1+0.4Z_3^3+0.2Z_4^{-4}-0.1Z_4^{-2}+0.3Z_4^0+0.1Z_4^2-0.1Z_4^4$ (Final). The theoretical values are the sum of the initial Zernike coefficients and the designed coefficients. The peak-valley (PV) values of the theoretical phase and the root-mean-square errors (RMSE) between the theoretical and final wavefront are given.}
\label{p4}
\end{figure*}

The modal method is by expanding an arbitrary phase function as $\phi(x,y)=\sum_{n=0}^{+\infty}\sum_{m}a_{n}^{m}Z_{n}^{m}(x,y)$, where $Z_{n}^{m}(x,y)$ are the Zernike polynomials \cite{Zernikepoly} defined on the unit disk, and $n,m$ are integers satisfying $n\geq0$ and $m=-n,-n+2,\dots,n$. We only consider $n\leq5$ here. Taking the partial derivatives with respect to $x$ and $y$, replacing differentials with differences for all $i$ points, and omitting the $Z_0^0(x,y)=1$ term, we have $2i$ equations. Using the least square method, we can obtain the 20 coefficients $a_n^m$ to reconstruct the wavefront. In our experiment, after calibrating SP with respect to the 880 nm light, we measured the phase gradient and obtained the coefficients using the 671 nm, 808 nm and 880 nm light before (Initial, the SLM is loaded only with the grating) and after (Final, the SLM is loaded with the grating and the designed phase aberration) modulation. The aberration is a superposition of Zernike polynomials from the second to the fourth order. The radius of the unit disk is set to be 750 \textmu m, so the following distributions are defined on $298\times298$ square meshes. The theoretical values are calculated by adding the designed coefficients to initial values. We calculated the peak-valley (PV) of the theoretical phase, and the RMSE from the theoretical to the final wavefront. The relative $k_x,k_y$ distributions, Zernike coefficients, phase distributions, and the calculated values are shown in Fig.~\ref{p4}. In spite of the errors in the coefficients, the theoretical and final phase distributions are similar. The wavefront correction task can be implemented using similar methods, by loading the opposite phase on the SLM.

The initial 671 nm and 808 nm wavefront are still oblique. Taking the largest initial first-order Zernike coefficient $a_1^1\approx0.690$ in 671 nm scenario as an example, we found the undesirable OPD is approximately $0.015\lambda$. To test whether they are from the misalignment of WMWS or the preparation process, we let the collimator directly face the WMWS, calibrated the SP with respect to 880 nm light, and switched the light source and QWP, to measure $a_1^{-1},a_1^1$ and calculate the corresponding OPD values as shown in Table.~\ref{zcopd}. There are remaining first-order coefficients in the 880 nm scenario as the calibration is not perfect, and the maximum OPD induced by switching the light source is roughly $0.006\lambda$, which means in our previous experiment, the SLM may be a source of angle deviation of the light beam, and the achromatism of our WMWS should be better.

\begin{table}[h]
\centering
\caption{\bf First-order Zernike Coefficients and Optical Path Differences of Different Light Sources}
\begin{tabular}{cccc}
\hline
Wavelength & 671 nm & 808 nm & 880 nm \\
\hline
$a_1^{-1}$ & $-0.1710$ & $-0.1767$ & $0.1156$ \\
$a_1^1$ & $0.2078$ & $0.2110$ & $-0.0207$ \\
OPD from $a_1^{-1}$ & $-0.0038\lambda$ & $-0.0038\lambda$ & $0.0025\lambda$ \\
OPD from $a_1^1$ & $-0.0046\lambda$ & $-0.0046\lambda$ & $0.0005\lambda$ \\
\hline
\end{tabular}
\label{zcopd}
\end{table}

In summary, we took advantage of the Savart plate and proposed a more realizable achromatic weak measurement wavefront sensor, which is also against lights with low temporal coherence. We discussed its spatial resolution, experimentally reconstructed a phase pattern to show it is higher than a common Shack–Hartmann sensor, and used the modal method to obtain the Zernike coefficients before and after phase modulation. Due to a lack of a superachromatic quarter-wave retarder, we still switched the QWP when changing $\lambda$. Even when we switch the QWP, after calibrating the SP with respect to one wavelength, a small phase difference still appears when switching to other $\lambda$ values, which is negligible in some cases, and can be eliminated by a slight recalibration. When using the original single-crystal WMWS, the OPD is often dozens of micrometers and the phase difference from changing $\lambda$ is large, and the recalibration requires more efforts. Our work is a new step in developing this new wavefront sensor without a delicate lenslet array. It can be applied when the sensitivity matters like the correction of atmospheric turbulence in astronomy \cite{Dayton:92} and the assessment of optical devices \cite{Jeong:05}, or when the requirement of spatial resolution is higher than the dynamic range like OPD measurement in biomedical imaging \cite{Gong:17} without microscopy. Also, it will be able to detect white lights when an achromatic SP \cite{Mu:14} and a superachromatic QWP are used.

\begin{backmatter}
\bmsection{Funding} Innovation Program for Quantum Science and Technology (2021ZD0301400); National Natural Science Foundation of China (61725504, U19A2075, 11821404); Anhui Initiative in Quantum Information Technologies (AHY020100, AHY060300); Fundamental Research Funds for the Central Universities (WK2030380017, WK5290000002).

\bmsection{Disclosures} The authors declare no conflicts of interest.

\bmsection{Data availability} Data underlying the results presented in this paper are not publicly available at this time but may be obtained from the authors upon reasonable request.

\end{backmatter}

\bibliography{sample}

\end{document}